\documentclass[]{aa}  
\usepackage{graphicx}
\usepackage{txfonts}
\usepackage{xcolor}
\usepackage{mathrsfs}
\newcommand{\bd}{\boldsymbol}
\newcommand{\vvec}{\bd{\varv}}
\newcommand{\dr}{\partial}
\newcommand{\gvec}{\vec{g}}
\newcommand{\Em}{\mathscr{E}}

\begin{document} 
   \title{ Effects of Radiation in Accretion Regions of Classical T Tauri Stars}
   \subtitle{Pre-heating of accretion column in non-LTE regime}

   \author{S. Colombo
          \inst{1,2,3}
          \and
          L. Ibgui\inst{2}
          \and
          S. Orlando \inst{3}
          \and
          R. Rodriguez \inst{5}
          \and
          G. Espinosa \inst{5}
          \and
          M. González \inst{4}
          \and
          C. Stehlé \inst{2}
          \and
          L. de S{\'a} \inst{2}
          \and
          C. Argiroffi \inst{1,3}
          \and
          R. Bonito \inst{3}
          \and
          G. Peres \inst{1,3}
          }

   \institute{Università degli Studi di Palermo, Dipartimento di Fisica e Chimica, via Archirafi 36, Palermo, Italy. \\
              \email{salvatore.colombo@inaf.it}
         \and
          LERMA, Observatoire de Paris, Sorbonne Université, Université de Cergy-Pontoise, CNRS, Paris, France.
          \and
          INAF - Osservatorio Astronomico di Palermo, Piazza del Parlamento 1, Palermo, Italy.
          \and
          Université Paris Diderot, Sorbonne Paris Cité, AIM, UMR7158, CEA, CNRS, F-91191 Gif-sur-Yvette, France 
          \and
          Universidad de Las Palmas de Gran Canaria, Gran Canaria, Spain. 
             }


 
  \abstract
 {Models and observations indicate that the impact of matter accreting onto the surface of young stars produces regions at the base of accretion columns, in which optically thin and thick plasma components coexist. Thus an accurate description of these impacts requires to account for the effects of absorption and emission of radiation.}
 {We study the effects of radiation emerging from shock-heated plasma in impact regions on the structure of the pre-shock downfalling material. We investigate if a significant absorption of radiation occurs and if it leads to a pre-shock heating of the accreting gas.}
 {We developed a radiation hydrodynamics model describing an accretion column impacting onto the surface of a Classical T Tauri Star. The model takes into account the stellar gravity, the thermal conduction, and the effects of both radiative losses and absorption of radiation by matter in the non local thermodynamic equilibrium regime.}
 {After the impact, a hot slab of post-shock plasma develops at the base of the accretion column. Part of radiation emerging from the slab is absorbed by the pre-shock accreting material. As a result, the pre-shock accretion column gradually heats up to temperatures of $10^5$~K, forming a radiative precursor of the shock. The precursor has a thermal structure with the hottest part at $T\approx 10^5$~K, with size comparable to that of the hot slab, above the post-shock region. At larger distances the temperature gradually decreases to $T \approx 10^4$~K.}
 {Our model predicts that $\approx 70$\% of radiation emitted by the post-shock plasma is absorbed by the pre-shock accretion column immediately above the slab and re-emitted in the UV band. This may explain why accretion rates derived from UV observations are systematically larger than rates inferred from X-ray observations.}

   \maketitle
%
\section{Introduction}
 Classical T Tauri stars (CTTSs) are young stars surrounded by a disk which extends internally up to the so called truncation radius \citep{2007prpl.conf..479B}. There, the magnetic field of the star is strong enough to govern the plasma dynamics: the plasma is funneled by the magnetic field into accretion columns, and falls onto the star \citep{1990RvMA....3..234C,1991Apj...370L..39K}. The material impacts the stellar surface producing shocks that dissipate the kinetic energy of the downfalling gas, thereby heating the material up to a few millions degrees. UV and X-ray observations of CTTSs show emission lines produced by dense material ($n \ge 10^{11}~$cm$^{-3}$), at temperature in the range $10^5 - 10^6$~K, which have been interpreted as originating from accretion impacts \citep{2007A&A...465L...5A,2007A&A...466.1111G}.

 Several models have been developed in order to study the impacts of accretion streams onto the surface of CTTSs, covering a wide range of physical effects and conditions (e.g. \citealt{2008A&A...491L..17S,2010A&A...522A..55S,2010A&A...510A..71O,2013A&A...559A.127O}). These models and the analysis of observations at different wavelengths show that impact regions can present complex structures and dynamics. The accretion columns themselves can be structured in density (\citealt{2011MNRAS.415.3380O, 2013A&A...557A..69M, 2014ApJ...795L..34B,2016A&A...594A..93C, 2019arXiv190207048C}), so that post-shock regions may show a wide range of densities, temperatures, and velocities. As a result, different plasma components, both optically thin and optically thick, and emitting in different wavelength bands, co-exist. In this scenario, reprocessing of radiation, through absorption and re-emission, is expected (\citealt{1998ApJ...509..802C,1998ARep...42..322L,2013Sci...341..251R, 2014ApJ...795L..34B, 2017SciA....3E0982R}) and may lead to complex profiles of emission lines.
 
Most of the models, however, do not take into account the effects of radiative absorption by matter, but only the radiative losses from optically thin plasma. The first attempt to include the full radiation effects in the context of accretion impacts was done by \cite{2017A&A...597A...1C}. In their model, the radiation is not fully coupled with hydrodynamic (HD) equations, but included in an iterative way. Nevertheless, their approach proved that the radiation emerging from the post-shock region is absorbed by the pre-shock accreting material, leading to a pre-shock heating of the accretion column (a radiative precursor). Since their approach is not fully self-consistent, these authors could not describe the structure of the precursor and derived a range of possible values of its temperature between $10^4 - 10^6$~K. More recently, \cite{de_Sa} described accretion impacts in CTTSs in the local thermodynamic equilibrium (LTE) regime and found that non-LTE is required for a correct description of the phenomenon.
 
 In this work, we investigate the effects of radiation on the structure of the accreting gas through a HD model of accretion impacts which includes self-consistently, for the first time, the effects of both radiative losses and absorption of radiation by matter in the non-LTE regime. Here we present the results for a typical accretion column which produces detectable X-ray emission at its impact (\citealt{2007A&A...465L...5A,2010A&A...522A..55S}).

%
\section{The radiation hydrodynamics model}

\begin{figure}[!t]
     \centering
     \includegraphics[scale = 0.12]{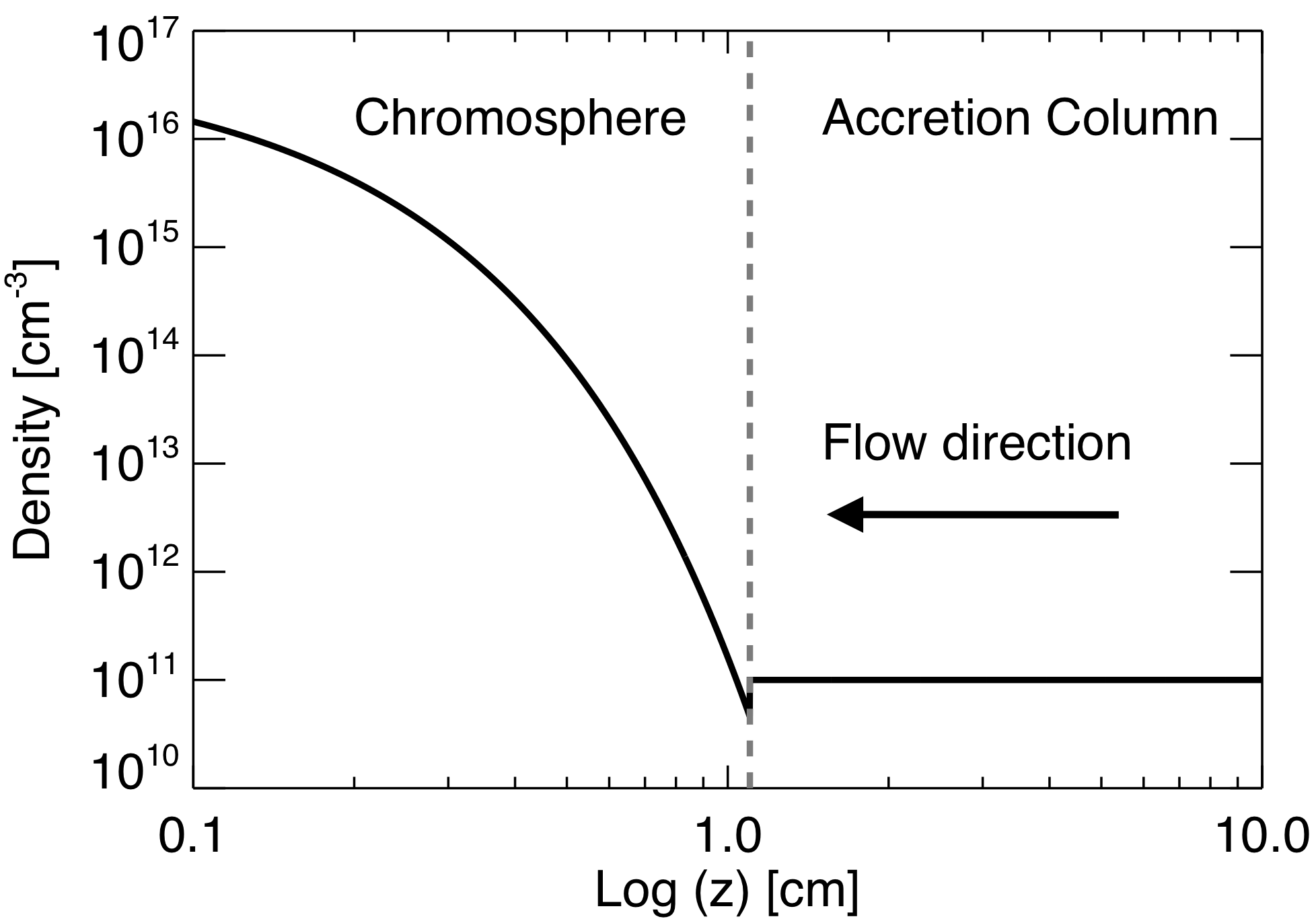}
     \caption{Initial profile of density along the z-axis (in Log scale). The dashed grey line separates the pre-shock chromosphere from the pre-shock accretion column. The arrow shows the flow direction.
     \label{img:ci}}
 \end{figure}

The model describes the impact of an accretion stream onto the chromosphere of a CTTS. The initial conditions are shown in Fig. \ref{img:ci}: they consist of a pre-shock accretion column with density $n = 10^{11}$~cm$^{-3}$ and temperature $T = 2 \times 10^4$~K\footnote{Since the temperature of the pre-shock accreting gas is not constrained by observations, here we adopt the minimum temperature for which the opacities are defined (in look-up tables; see later in the text). Nevertheless, in photoionized plasmas, we expect that the opacities at lower temperatures (down to $10^3~$K) are comparable with those at $T = 2 \times 10^4$~K \citep{2009A&A...508.1539M,2013MNRAS.429.3133L}.}, and the  pre-shock chromosphere assumed to be isothermal with temperature $T = 10^4$~K. Initially, the accreting pre-shock plasma is in radiative equilibrium; it flows along the $z$-axis and at time = $0$ it impacts onto the chromosphere. We expect that, after the impact, a shock will be generated at the interface between the accretion column and the chromosphere; furthermore, a reverse shock will propagate along the accretion column.


 The model takes into account the gravity field $\vec{g}$ of the star, the thermal conduction $\vec{F_c}$ (including the classical and saturated regime as described in \citealt{2016A&A...594A..93C}), the radiative losses $L$ and the radiative gains. The model solves the time-dependent equations of conservation of mass, momentum and energy using the perfect gas equation of state
 
\begin{equation}
     \frac{\dr \rho}{\dr t} + \nabla \cdot \left( \rho \, \vvec\right) = 0
    \label{mass}
\end{equation}

\begin{equation}
   \frac{\dr}{\dr t}\left( \rho \, \vvec \right) +  \nabla \cdot \left( \rho \, \vvec \otimes \vvec + p \, \mathbb{I} \right) = \rho \, \gvec + \frac{\rho \, k_R}{c} \vec{F}
    \label{eq:01}
\end{equation}
    
\begin{equation}
      \begin{alignedat}{2}
   &\frac{\dr}{\dr t} \left( \Em + \tfrac{1}{2} \, \rho \, \varv^2 \right) +  \nabla \cdot \left[ \left( \Em + \tfrac{1}{2} \, \rho \, \varv^2 + p \right) \, \vvec  \right] = \rho \, \gvec \cdot \vvec \\
   & \hspace{2cm}
    - \nabla \cdot \vec{F_c} + \frac{\rho \, k_R}{c} \, \vec{F} \cdot \vvec + \, c \, \rho \, k_P  \, E - L
    \end{alignedat}
    \label{eq:energy_cons}
\end{equation}
    
\begin{equation}
    \frac{\partial E}{\partial t} + \vec{\nabla}\cdot \vec{F} \, = \, L - c \, \rho \, k_P \, E \label{energy}
\end{equation}

\begin{equation}
    \vec{F} \, = \, - \left(\frac{c \, \lambda}{\rho \, k_R}\right) \, \nabla E
\end{equation}

\begin{equation}
     p = \rho \, \frac{k_B \, T}{\mu \, m_H}
\end{equation}

\begin{equation}    
   \Em = \frac{p}{\gamma -1}
\end{equation}

\noindent
 where $\rho$ is the plasma mass density,  $\vvec$ is the fluid velocity, $k_R$ and $k_P$ are, respectively, the Rosseland and Planck mean opacities, $c$ the speed of light, $\vec{F}$ the comoving-frame radiation flux,  $\Em$ the plasma internal energy density, $E$ the comoving-frame radiation energy, $k_B$ the Boltzmann constant, $T$ the plasma temperature, $\mu$ the mean particle weight (assuming solar abundances), $m_H$ the hydrogen mass, $\lambda$ the flux-limiter \citep{MINERBO1978541}, and $\gamma = 5/3$ the ratio of specific heats. In a companion paper (\citealt{Colombo_et_al_2019_2}; in the following Paper I), we discuss the assumptions and the limits of these equations. 
 Since we are interested in evaluating the effects of irradiation from the post-shock accreting material on the pre-shock accretion column, we neglect the radiation effects (both radiative losses and absorption of radiation by matter) in the pre-shock chromosphere (i.e., $k_P = k_R =0$ and $L =0$ in Eqs. 1-4).

 

 The calculations are performed using PLUTO \citep{2007ApJS..170..228M}, a modular, Godunov-type parallel code for astrophysical plasma. The HD equations are solved using the HD module available in PLUTO for the linearized Roe Riemann solver based on characteristic decomposition of the Roe matrix \citep{ROE}. The time evolution is solved using a second order Runge-Kutta method. The thermal conduction is treated with the super-time-stepping method \citep{CNM:CNM950}. 

 The radiation effects are calculated by coupling PLUTO with a radiation module which was originally restrained to the LTE regime \citep{2013A&A...559A..80K}, and which we have extended in order to take into account the non-LTE conditions. 
 In Paper I, we describe the details of the implementation and the limits and assumptions of the new numerical module (see also Appendix A). The radiation quantities ($k_P$, $k_R$, $L$) are calculated at the temperature and density of interest using look-up tables \citep{PhysRevE.98.033213} and a bi-linear interpolation method. The radiation module makes use of the "Portable, Extensible Toolkit for Scientific Computation" (PETSc) library to solve the radiative part of the system \citep{petsc-user-ref}. In particular, the PETSc library uses the Krylov subspace iterative method and a preconditioner to solve the matrix equation. For this work, the generalized minimal residual as iterative method and block Jacobi as preconditioner are used.

\begin{figure}[!t]
     \centering
     \includegraphics[scale = 0.5]{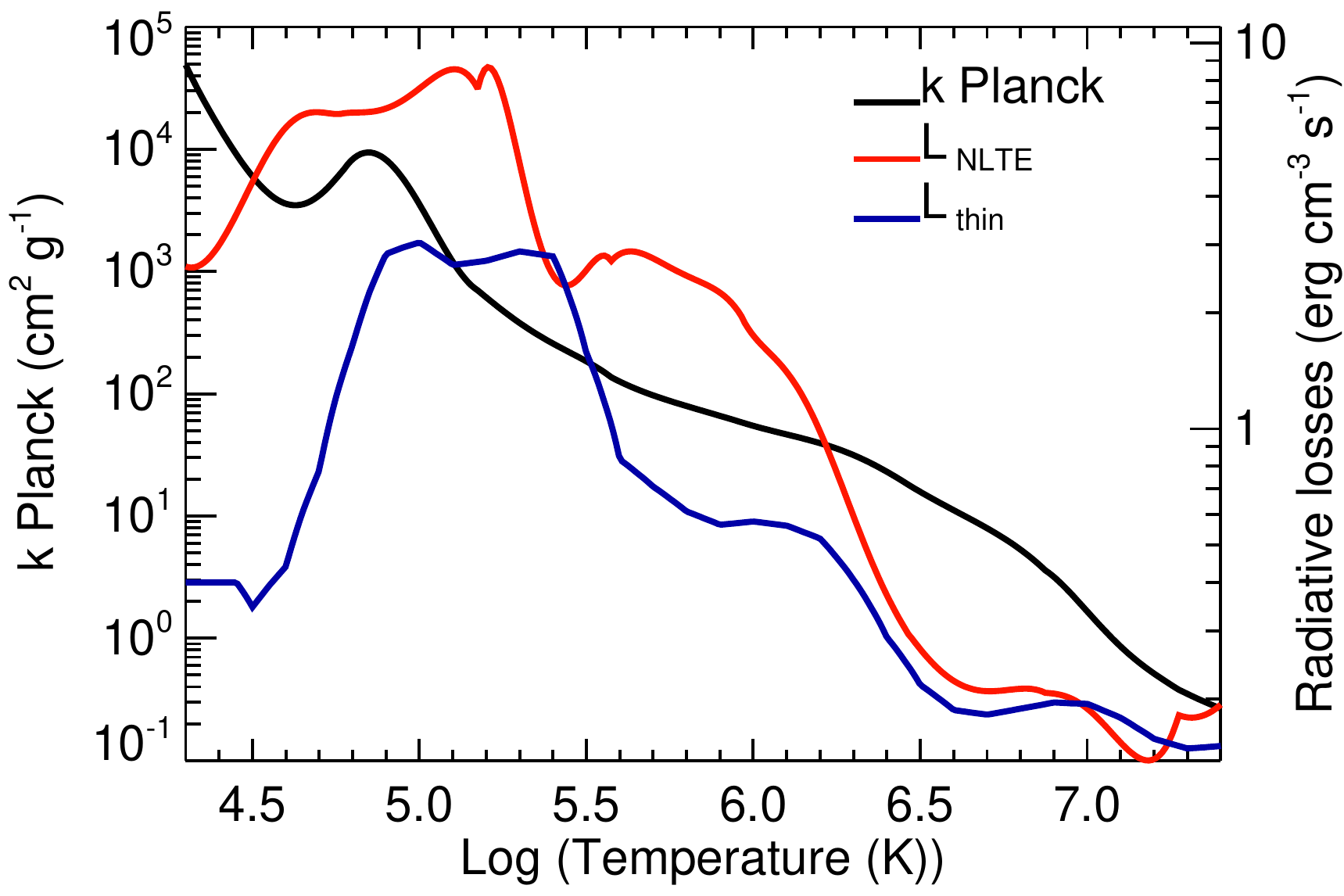}
     \caption{Planck opacity ($k_P$, black) and radiative losses in non-LTE ($L_{NLTE}$, red) at the density of the accretion column (i.e., $n = 10^{11}~$~cm$^{-3}$) versus temperature. For comparison, the figure shows also the radiative losses from optically thin plasma ($L_{thin}$, blue) used in models available in literature (e.g. \citealt{2008A&A...491L..17S}).
     \label{img:op}}
 \end{figure}

 We considered the non-LTE regime and used the frequency integrated approach, in which the radiation quantities $k_P$, $k_R$ and $L$ and radiation energy equation are integrated over all frequencies from 0 to $\infty$ (\citealt{PhysRevE.98.033213}). As an example, Fig. \ref{img:op} shows the value of $k_P$ and $L$ calculated at the density of the pre-shock accretion column (namely $n = 10^{11}~$~cm$^{-3}$). In Sect.~3.2, we discuss the limits of the frequency integrated approach. 

 The radiation module works only in 3D \citep{2013A&A...559A..80K}. To reduce the computational cost, we assume that the gas flows along the $z$-axis and that the domain consists of a 3D Cartesian grid with 3 cells for the $x$- and $y$-axes and 9048 cells for the $z$-axis (so the model can be considered as 1D). In other words, the model describes the flow dynamics along the axis of an accretion stream. The $x$- and $y$-axes are both covered with a uniform grid extending for $1.4\times10^7$cm. Along the $z$-axis, the grid is non-uniform and composed of 2 patches. The first covers the domain between $10^8$~cm and $2\times 10^{9}$~cm (thus fully including the chromosphere which extends up to $1.1\times 10^9$~cm) with a uniform grid with spatial resolution of $\approx 9.7\times10^5$~cm (2048 cells); the second patch covers the domain between $2\times 10^{9}$~cm and $10^{10}$~cm with a stretched grid, with a minimum spatial resolution of $\approx 9.7\times10^5$~cm close to the first patch, and a maximum spatial resolution of $\approx 13\times 10^5$~cm at the end of the domain (7000 cells). The choice of this grid provides the best compromise between accuracy and efficiency of the calculation and allows to treat accurately the effects of radiative cooling instability in the post-shock region. The bottom ($-z$) and top ($+z$) boundaries are fixed: the first to values consistent with the chromosphere and the second with values describing an inflow of material with constant density, velocity and radiation energy, the latter chosen in such a way that the material is in radiative equilibrium with it.
 We set the lateral boundaries as periodic.
 
 We present two simulations. The first (run RHD) describes the impact of the accretion column, taking into account all the effects described above. The second (run HD) describes the same impact, but in the optically thin regime, namely using the radiative losses adopted by \cite{2008A&A...491L..17S} (see Fig.~\ref{img:op}) and without radiative absorption; this simulation produces results analogous to those of previous models \citep[e.g.][]{2008A&A...491L..17S,2010A&A...522A..55S}, and it is used for comparison with run RHD to highlight the effects of absorption. 
%
\section{Results}
\subsection{Dynamics of the post-shock plasma\label{dynamics}}

\begin{figure}[!t]
    \centering
    \includegraphics[scale = 0.5]{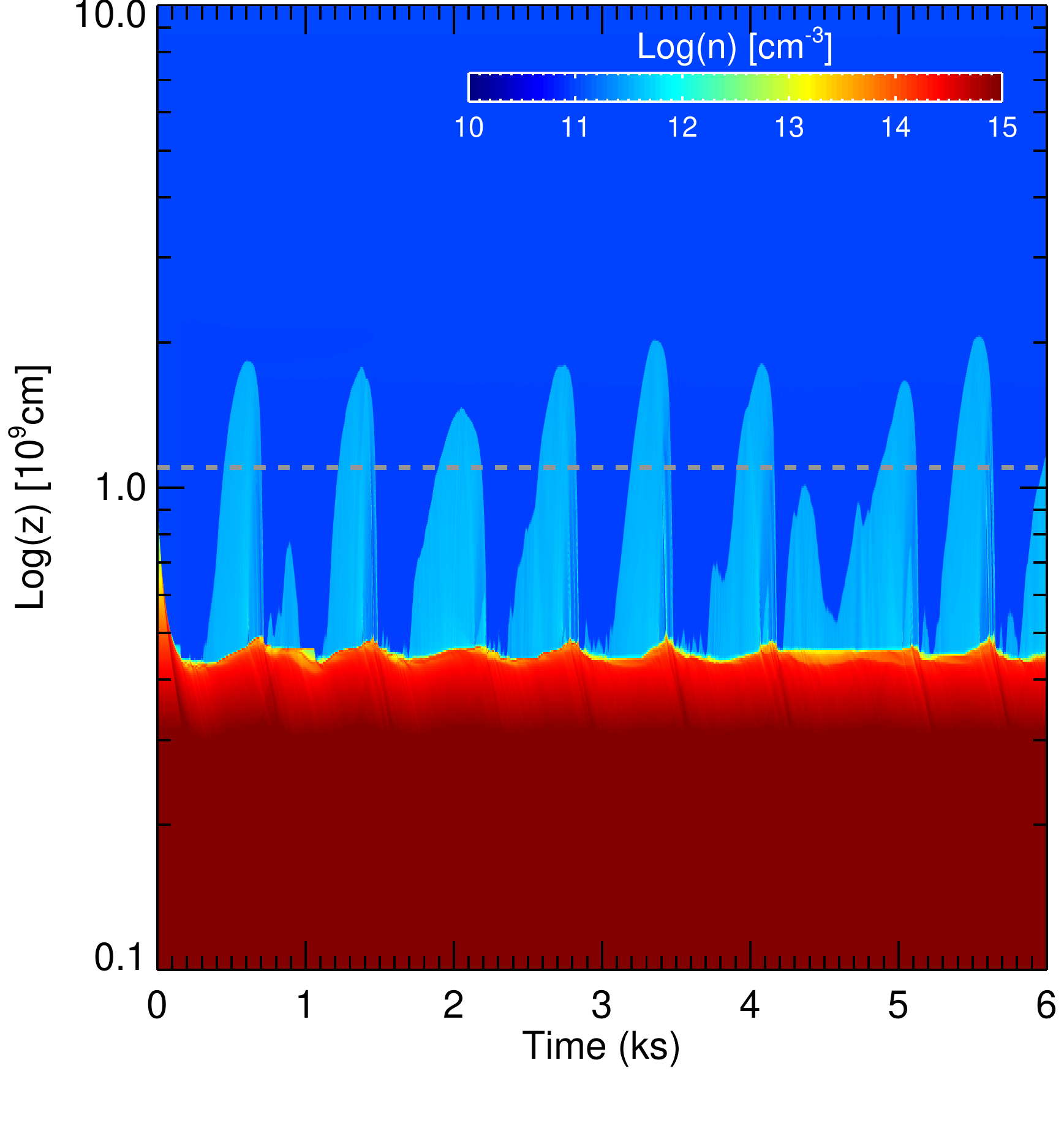}
    \centering
    \includegraphics[scale = 0.5]{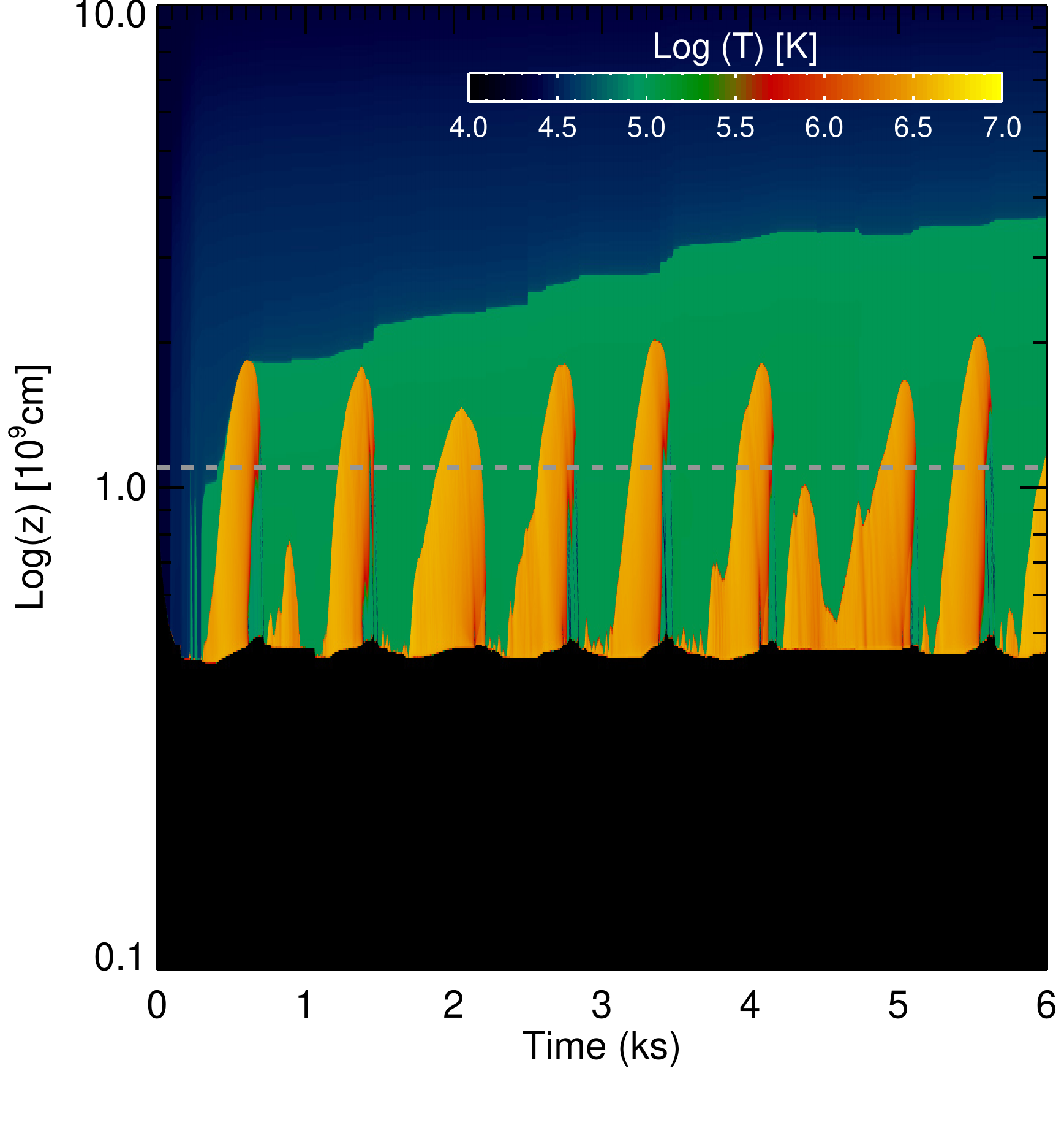}
    \caption{Space-time maps, in logarithmic scale, of density (top panel) and temperature (bottom panel) for run RHD. The green region in the bottom panel corresponds to the hottest part of the precursor. The grey dotted lines in both panels mark the pre-impact position of the chromosphere.}
    \label{img:EV}
\end{figure}

 We follow the evolution of the system for $\approx 6$~ks. The dynamics is similar to that of models describing accretion impacts without radiation effects  in the context of both CTTSs (e.g. \citealt{2008A&A...491L..17S,2010A&A...522A..55S}), and magnetic cataclysmic variables (e.g. \citealt{2015A&A...579A..25B,2017A&A...600A..53M,2018MNRAS.473.3158V}). Fig. \ref{img:EV} shows the space-time maps of density and temperature for the whole evolution of run RHD. Initially the accretion column hits the chromosphere with a speed of 500~km~s$^{-1}$ and sinks deeply onto the chromosphere, down to the position at which the ram pressure of the downflowing material equals the thermal pressure of the chromosphere. Then a shock develops at the base of the accretion column. The dynamics of the post-shock plasma is analogous in the two runs considered.
 The shock propagates upward through the accretion column (expanding phase), heating the plasma up to temperatures of a few million degrees and generating a hot and dense post-shock region (slab) that extends up to $\approx 2\times 10^9~$cm (light blue regions in the top panel and yellow to orange regions in the bottom panel of Fig. \ref{img:EV}). During the expansion of the post-shock, the density increases at the base of the slab due to accumulation of the accreting material. As a consequence, the radiative losses in that region rise. The expanding phase ends when the radiative losses in the slab reach a critical value that triggers the formation of cooling structures due to thermal instability. The plasma rapidly cools down at the base of the slab, its pressure decreases, and the slab is not able to sustain the material above. As a result, the region collapses (collapse phase). After the collapse, a new shock forms due to the continuous downflow and a new slab expands until it collapses under the action of radiative losses. The result of this evolution is shown in Fig. \ref{img:EV}, where the alternating phases of expansion and collapse of the post-shock region are clearly visible. It is worth noting that the quasi-periodic oscillations of the post-shock region have never been observed in CTTSs. In fact, accretion streams are expected to be structured in several fibrils, each independent of the others due to the strong magnetic field which prevents mass and energy exchange across magnetic field lines. The fibrils can be characterized by slightly different densities, which would result in different instability periods, as also by random phases of oscillations. Thus, for a stream composed of several fibrils, the integrated emission is expected to show no evident periodic modulation \citep{2010A&A...510A..71O,2013A&A...557A..69M,2017A&A...600A..53M}.
 
 \subsection{Effects of radiation}
 
\begin{figure}
     \centering
     \includegraphics[scale = 0.6]{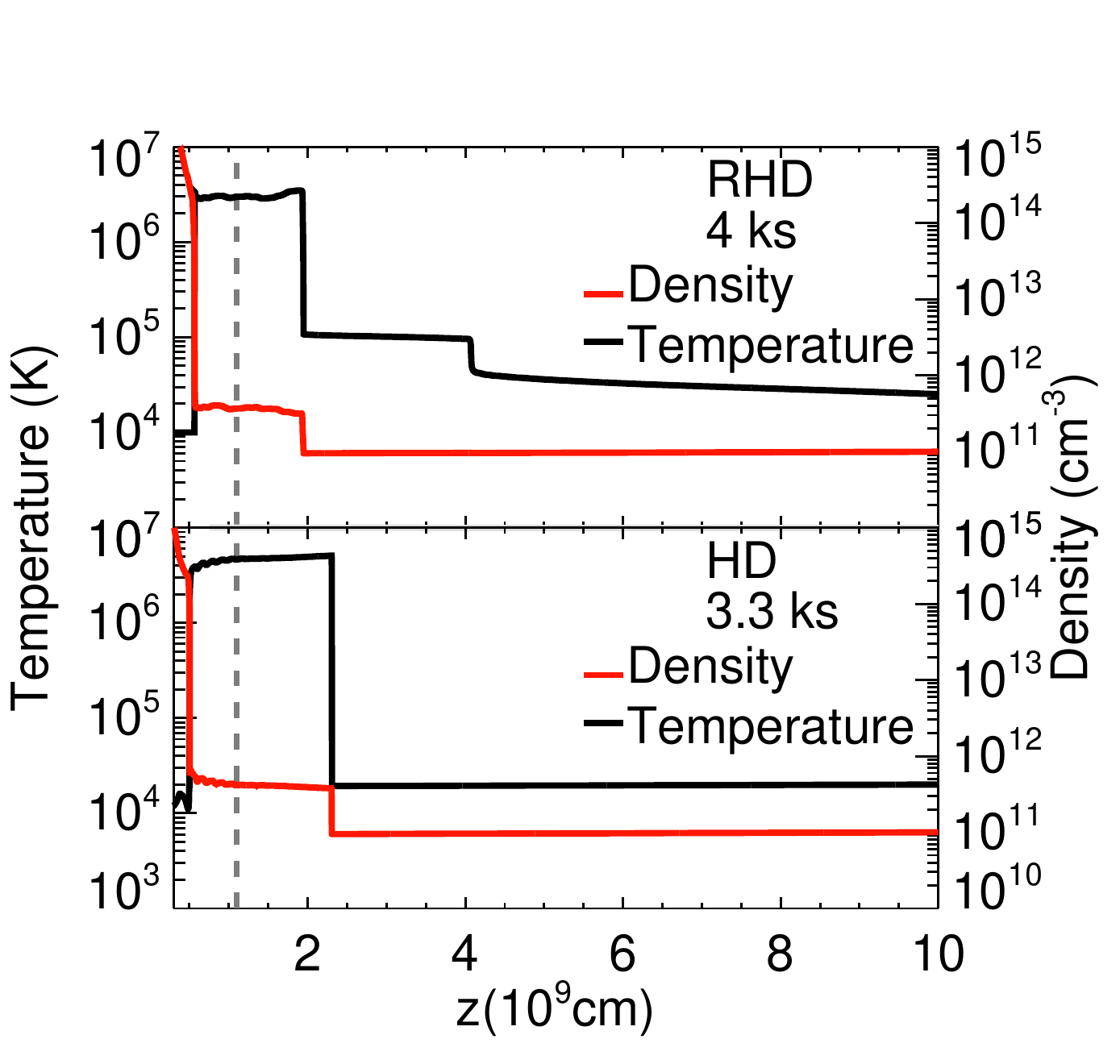}
     \caption{Temperature (black line) and density (red line) profiles for runs RHD (top panel) and HD (bottom panel), during one of the expanding phases. The grey dotted line marks the initial position of the chromosphere.}
     \label{img:profile}
 \end{figure}

 The effects of radiation are evident in the pre-shock downflowing plasma. Fig. \ref{img:profile} shows the vertical profiles of temperature and density for the two runs considered, during one of the expansion phases. The temperature profile in run RHD presents two main features: the hot post-shock with $T\approx 3$~MK located between $z = 0.5\times10^{9}~$cm and $z = 2\times10^{9}~$cm (present also in run HD and in previous HD models with no radiation effects included), and a precursor region with temperature ranging between $2\times 10^4$~K and $10^5$~K (up to $z = 10^{10}$~cm). The radiative precursor of the shock is not present in run HD (see bottom panel in Fig.~\ref{img:profile}) nor in previous HD models. We note that thermal conduction cannot contribute in producing the precursor in our simulations due to heat flux saturation effects at the shock front (\citealt{2005A&A...444..505O}).
 In the density profile, the post-shock slab with a density a factor of 4 larger than that of the pre-shock material is clearly visible (as expected from the strong shock conditions \citealt{1967pswh.book.....Z}).
  
 The precursor is the result of irradiation of the pre-shock material by the post-shock plasma. Since the pre-shock accreting material is optically thick, we found that it absorbs $\approx 70 \%$ of radiation immediately above the slab at an height of $z = 4\times 10^9~$cm from the chromosphere and heats up to maximum temperatures around $10^5$~K (see Fig.~\ref{img:EV}). 

 We observe that the precursor is structured in temperature. After the early transient phase, i.e, when the hot part of the precursor does not evolve any more ($t> 3.5$~ks), the whole precursor in run RHD extends up to the limit of the domain at $z = 10^{10}~$cm. During the downflow, the accreting material (with an initial temperature of $T= 2\times 10^4$~K) absorbs part of the radiation from the slab and, as a result, is gradually heated up to $T\approx 6\times 10^4$~K at $z \approx 4\times 10^{9}~$cm. At this point, an additional heating determines a sudden increase of temperature up to values around $10^5$~K (see upper panel of Fig.~\ref{img:profile}). This is due to a peak in absorption of radiation by matter around $T \approx 7\times 10^4$~K.
 In fact, at this temperature, the absorbed radiation energy, defined as $(k_P \rho c E)$ (see Eq. \ref{eq:energy_cons}) has a maximum due to the peak in $k_P$ (see Fig. \ref{img:op}), which generates the increase in the gas temperature.
 After the heating at $z \approx 4\times 10^{9}~$cm, the accreting material continues to fall with temperature around $10^5~$K until it is shocked. We note that, at this temperature, the radiative losses have a maximum (see Fig. \ref{img:op}) that prevents the material to heat up more. The peak in absorption around $T = 7\times10^4~$K is due to helium (see \citealt{PhysRevE.98.033213}) which, therefore, plays a critical role in determining the structure of the radiative precursor (see Appendix B).
 

 \subsection{Distribution of Emission Measure versus Temperature}
 
 \begin{figure}[!t]
    \centering
    \includegraphics[scale = 0.5]{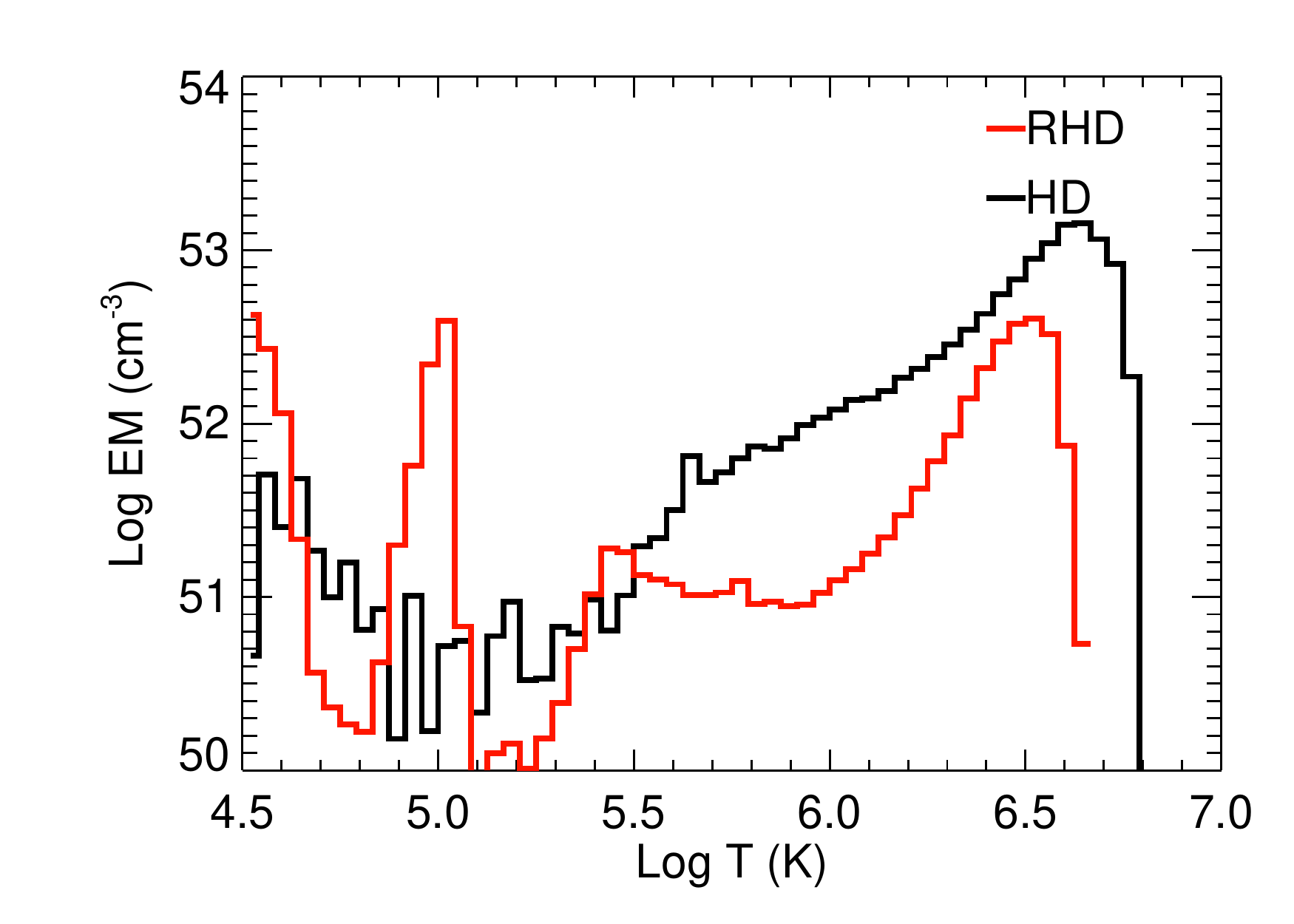}
    \caption{EM synthesized from runs RHD (red histogram) and HD (black) versus temperature.}
    \label{img:EM}
\end{figure}

From the models, we derive the time-averaged distributions of emission measure versus temperature ($EM = \int_V n^2 dV$) which is useful to obtain information about the plasma components emitting at various temperatures. The distributions are derived by following the approach described by \cite{2016A&A...594A..93C}, assuming a stream with a mass accretion rate of $10^{-9.17}\,M_{\odot}$~yr$^{-1}$ as for TW Hya (e.g. \citealt{2011A&A...526A.104C}). This would require a column with a reasonable radius $r_s \approx 3.5 \times 10^{10}$ cm. Fig.~\ref{img:EM} compares the distributions synthesized from runs RHD and HD to highlight the effects of radiation. The most striking difference between the two distributions is that the EM of RHD (red histogram in Fig. \ref{img:EM}) shows three main peaks at $\log T= 4.5$, $5.0$ and $6.5$, whereas the EM of HD (black histogram in Fig. \ref{img:EM}) has a prominent peak at $\log T= 6.6$ and a small peak at $\log T = 4.7$. The origin of the peaks in the EM of run RHD can be investigated by considering the bottom panel of Fig. \ref{img:EV}. The orange regions in the figure mark the plasma with $\log T> 6$, belonging to the hot slab. Thus, the highest temperature peak in the two models considered is clearly due to the post-shock plasma of the slab. This peak is shifted at slightly lower temperatures and lower emission measure values in run RHD due to the radiative losses which are more efficient in run RHD than in HD (see Fig.~\ref{img:op}).

The other two peaks at lower temperatures in the EM distribution of run RHD originate from blue and green regions in the bottom panel of Fig. \ref{img:EV} which mark the plasma with $4.5 < \log T < 4.8$ and $4.8 < \log T < 5.3$, respectively. These regions identify the precursor in run RHD: the peak at $\log T \approx 5$ in the EM distribution corresponds to the hottest part of the precursor. This feature is not present in run HD. 

In addition, the EM distributions have a different slope in the region between $5.5<\log T<6.0$, again due to the differences in the radiative losses adopted in the two models (see Fig. \ref{img:op}). The EM of RHD shows also a small bump between $\log T = 5.3$ and $\log T = 5.9$ not present in run HD. This feature originates from the plasma cooling during the collapse phase (red regions in Fig.\ref{img:EV}) and is due to absorption of radiation by this plasma component. 
\section{Conclusions}

We have developed a model which describes the radiation effects in the non-LTE regime during the impact of an accretion column onto the surface of a CTTS. Our model shows that a significant fraction ($\approx 70 \%$ at a distance of $4\times 10^9~$cm from the chromosphere) of radiation produced in the post-shock plasma is absorbed by the pre-shock material of the accretion column. As a result, this material is heated up to temperature of the order of $10^5$~K, forming a radiative precursor. The precursor reaches the highest temperature in proximity of the shock. At $z\approx 4\times 10^9~$cm, the temperature of the precursor decreases to temperatures between $2\times 10^4$~K and $5\times 10^4$~K. The precursor is a strong source of UV emission. The partial absorption of X-ray emission from the slab and the contribution of UV emission from the precursor may explain why accretion rates derived from UV observations are systematically larger than accretion rates inferred from X-ray observations (\citealt{2011A&A...526A.104C}). In future works, we will extend the analysis by exploring the parameter space of the model with the aim of investigating how the structure and evolution of the radiative precursor depend on the physical conditions (density, downflow velocity, abundance) of the accretion stream.

\section*{Acknowledgements}
We thanks the second anonymous referee that, with their expertise in the field, gave us helpful comments and suggestions that helped us to improved the manuscript.
We are grateful to Mario Floch for his help and the fruitful scientific discussions. 
PLUTO is developed at the Turin Astronomical Observatory and the Department of Physics of Turin University. We acknowledge the "Accordo Quadro INAF-CINECA (2017)”, the CINECA Award HP10B1GLGV and the HPC facility (SCAN) of the INAF – Osservatorio Astronomico di Palermo, for the availability of high performance computing resources and support. This work was supported by the Programme National de Physique Stellaire (PNPS) of CNRS/INSU co-funded by CEA and CNES. This work has been done within the LABEX Plas@par project, and received financial state aid managed by the Agence Nationale de la Recherche (ANR), as part of the programme "Investissements d'avenir" under the reference ANR-11-IDEX-0004-02. SO, RB, and CA acknowledge financial contribution from the agreement ASI-INAF n.2017-14.H.O.

\bibliographystyle{aa} 
\bibliography{Colombo}

\begin{thebibliography}{38}
\expandafter\ifx\csname natexlab\endcsname\relax\def\natexlab#1{#1}\fi

\bibitem[{{Alexiades} {et~al.}(1996){Alexiades}, {Amiez}, \&
  {Gremaud}}]{CNM:CNM950}
{Alexiades}, V., {Amiez}, G., \& {Gremaud}, P.~A. 1996, Communications in
  Numerical Methods in Engineering, 12, 31

\bibitem[{{Argiroffi} {et~al.}(2007){Argiroffi}, {Maggio}, \&
  {Peres}}]{2007A&A...465L...5A}
{Argiroffi}, C., {Maggio}, A., \& {Peres}, G. 2007, \aap, 465, L5

\bibitem[{Balay {et~al.}(2012)Balay, Abhyankar, Adams, Brown, Brune,
  Buschelman, Dalcin, Dener, Eijkhout, Gropp, Kaushik, Knepley, May, McInnes,
  Mills, Munson, Rupp, Sanan, Smith, Zampini, Zhang, \& Zhang}]{petsc-user-ref}
Balay, S., Abhyankar, S., Adams, M.~F., {et~al.} 2012, {PETS}c Users Manual,
  Tech. Rep. ANL-95/11 - Revision 3.3, Argonne National Laboratory

\bibitem[{{Bonito} {et~al.}(2014){Bonito}, {Orlando}, {Argiroffi}, {Miceli},
  {Peres}, {Matsakos}, {Stehle}, \& {Ibgui}}]{2014ApJ...795L..34B}
{Bonito}, R., {Orlando}, S., {Argiroffi}, C., {et~al.} 2014, \apjl, 795, L34

\bibitem[{{Bouvier} {et~al.}(2007){Bouvier}, {Alencar}, {Harries},
  {Johns-Krull}, \& {Romanova}}]{2007prpl.conf..479B}
{Bouvier}, J., {Alencar}, S.~H.~P., {Harries}, T.~J., {Johns-Krull}, C.~M., \&
  {Romanova}, M.~M. 2007, Protostars and Planets V, 479

\bibitem[{{Busschaert} {et~al.}(2015){Busschaert}, {Falize}, {Michaut},
  {Bonnet-Bidaud}, \& {Mouchet}}]{2015A&A...579A..25B}
{Busschaert}, C., {Falize}, {\'E}., {Michaut}, C., {Bonnet-Bidaud}, J.~M., \&
  {Mouchet}, M. 2015, \aap, 579, A25

\bibitem[{{Calvet} \& {Gullbring}(1998)}]{1998ApJ...509..802C}
{Calvet}, N. \& {Gullbring}, E. 1998, \apj, 509, 802

\bibitem[{{Camenzind}(1990)}]{1990RvMA....3..234C}
{Camenzind}, M. 1990, in Reviews in Modern Astronomy, Vol.~3, Reviews in Modern
  Astronomy, ed. G.~{Klare}, 234--265

\bibitem[{{Colombo} {et~al.}(2019{\natexlab{a}}){Colombo}, {Ibgui}, {Orlando},
  {Rodriguez}, {Espinosa}, {Gonz{\'a}lez}, {Stehl{\'e}}, {de S{\`a}},
  {Argiroffi}, {Bonito}, \& {Peres}}]{Colombo_et_al_2019_2}
{Colombo}, S., {Ibgui}, L., {Orlando}, S., {et~al.} 2019{\natexlab{a}}, \aap,
  {in press}, ArXiv http://arxiv.org/abs/1907.04591

\bibitem[{{Colombo} {et~al.}(2016){Colombo}, {Orlando}, {Peres}, {Argiroffi},
  \& {Reale}}]{2016A&A...594A..93C}
{Colombo}, S., {Orlando}, S., {Peres}, G., {Argiroffi}, C., \& {Reale}, F.
  2016, \aap, 594, A93

\bibitem[{{Colombo} {et~al.}(2019{\natexlab{b}}){Colombo}, {Orlando, S.},
  {Peres, G.}, {Reale, F.}, {Argiroffi, C.}, {Bonito, R.}, {Ibgui, L.}, \&
  {Stehl\'e, C.}}]{2019arXiv190207048C}
{Colombo}, S., {Orlando, S.}, {Peres, G.}, {et~al.} 2019{\natexlab{b}}, A\&A,
  624, A50

\bibitem[{{Costa} {et~al.}(2017){Costa}, {Orlando}, {Peres}, {Argiroffi}, \&
  {Bonito}}]{2017A&A...597A...1C}
{Costa}, G., {Orlando}, S., {Peres}, G., {Argiroffi}, C., \& {Bonito}, R. 2017,
  \aap, 597, A1

\bibitem[{{Curran} {et~al.}(2011){Curran}, {Argiroffi}, {Sacco}, {Orlando},
  {Peres}, {Reale}, \& {Maggio}}]{2011A&A...526A.104C}
{Curran}, R.~L., {Argiroffi}, C., {Sacco}, G.~G., {et~al.} 2011, \aap, 526,
  A104

\bibitem[{{de S{\'a}} {et~al.}(2019){de S{\'a}}, {Chi{\`e}ze}, {Stehl{\'e}},
  {Hubeny}, {Lanz}, {Cayatte}, \& {Delahaye}}]{de_Sa}
{de S{\'a}}, L., {Chi{\`e}ze}, J.-P., {Stehl{\'e}}, C., {et~al.} 2019, arXiv
  e-prints, arXiv:1904.09156

\bibitem[{{Gonz{\'a}lez} {et~al.}(2007){Gonz{\'a}lez}, {Audit}, \&
  {Huynh}}]{Gonzalez_et_al_2007}
{Gonz{\'a}lez}, M., {Audit}, E., \& {Huynh}, P. 2007, \aap, 464, 429

\bibitem[{{G{\"u}nther} {et~al.}(2007){G{\"u}nther}, {Schmitt}, {Robrade}, \&
  {Liefke}}]{2007A&A...466.1111G}
{G{\"u}nther}, H.~M., {Schmitt}, J.~H.~M.~M., {Robrade}, J., \& {Liefke}, C.
  2007, \aap, 466, 1111

\bibitem[{{Hayes} \& {Norman}(2003)}]{Hayes_and_Norman_2003}
{Hayes}, J.~C. \& {Norman}, M.~L. 2003, \apjs, 147, 197

\bibitem[{{Koenigl}(1991)}]{1991Apj...370L..39K}
{Koenigl}, A. 1991, \apj, 370, L39

\bibitem[{{Kolb} {et~al.}(2013){Kolb}, {Stute}, {Kley}, \&
  {Mignone}}]{2013A&A...559A..80K}
{Kolb}, S.~M., {Stute}, M., {Kley}, W., \& {Mignone}, A. 2013, \aap, 559, A80

\bibitem[{{Lamzin}(1998)}]{1998ARep...42..322L}
{Lamzin}, S.~A. 1998, Astronomy Reports, 42, 322

\bibitem[{{Lykins} {et~al.}(2013){Lykins}, {Ferland}, {Porter}, {van Hoof},
  {Williams}, \& {Gnat}}]{2013MNRAS.429.3133L}
{Lykins}, M.~L., {Ferland}, G.~J., {Porter}, R.~L., {et~al.} 2013, \mnras, 429,
  3133

\bibitem[{{Marigo} \& {Aringer}(2009)}]{2009A&A...508.1539M}
{Marigo}, P. \& {Aringer}, B. 2009, \aap, 508, 1539

\bibitem[{{Matsakos} {et~al.}(2013){Matsakos}, {Chi{\`e}ze}, {Stehl{\'e}},
  {Gonz{\'a}lez}, {Ibgui}, {de S{\'a}}, {Lanz}, {Orlando}, {Bonito},
  {Argiroffi}, {Reale}, \& {Peres}}]{2013A&A...557A..69M}
{Matsakos}, T., {Chi{\`e}ze}, J.-P., {Stehl{\'e}}, C., {et~al.} 2013, \aap,
  557, A69

\bibitem[{{Mignone} {et~al.}(2007){Mignone}, {Bodo}, {Massaglia}, {Matsakos},
  {Tesileanu}, {Zanni}, \& {Ferrari}}]{2007ApJS..170..228M}
{Mignone}, A., {Bodo}, G., {Massaglia}, S., {et~al.} 2007, \apjs, 170, 228

\bibitem[{Minerbo(1978)}]{MINERBO1978541}
Minerbo, G.~N. 1978, Journal of Quantitative Spectroscopy and Radiative
  Transfer, 20, 541

\bibitem[{{Mouchet} {et~al.}(2017){Mouchet}, {Bonnet-Bidaud}, {Van Box Som},
  {Falize}, {Buckley}, {Breytenbach}, {Ashley}, {Marsh}, \&
  {Dhillon}}]{2017A&A...600A..53M}
{Mouchet}, M., {Bonnet-Bidaud}, J.~M., {Van Box Som}, L., {et~al.} 2017, \aap,
  600, A53

\bibitem[{{Orlando} {et~al.}(2013){Orlando}, {Bonito}, {Argiroffi}, {Reale},
  {Peres}, {Miceli}, {Matsakos}, {Stehl{\'e}}, {Ibgui}, {de Sa}, {Chi{\`e}ze},
  \& {Lanz}}]{2013A&A...559A.127O}
{Orlando}, S., {Bonito}, R., {Argiroffi}, C., {et~al.} 2013, \aap, 559, A127

\bibitem[{{Orlando} {et~al.}(2005){Orlando}, {Peres}, {Reale}, {Bocchino},
  {Rosner}, {Plewa}, \& {Siegel}}]{2005A&A...444..505O}
{Orlando}, S., {Peres}, G., {Reale}, F., {et~al.} 2005, \aap, 444, 505

\bibitem[{{Orlando} {et~al.}(2011){Orlando}, {Reale}, {Peres}, \&
  {Mignone}}]{2011MNRAS.415.3380O}
{Orlando}, S., {Reale}, F., {Peres}, G., \& {Mignone}, A. 2011, \mnras, 415,
  3380

\bibitem[{{Orlando} {et~al.}(2010){Orlando}, {Sacco}, {Argiroffi}, {Reale},
  {Peres}, \& {Maggio}}]{2010A&A...510A..71O}
{Orlando}, S., {Sacco}, G.~G., {Argiroffi}, C., {et~al.} 2010, \aap, 510, A71

\bibitem[{{Reale} {et~al.}(2013){Reale}, {Orlando}, {Testa}, {Peres}, {Landi},
  \& {Schrijver}}]{2013Sci...341..251R}
{Reale}, F., {Orlando}, S., {Testa}, P., {et~al.} 2013, Science, 341, 251

\bibitem[{Revet {et~al.}(2017)Revet, Chen, Bonito, Khiar, Filippov, Argiroffi,
  Higginson, Orlando, B{\'e}ard, Blecher, Borghesi, Burdonov, Khaghani,
  Naughton, P{\'e}pin, Portugall, Riquier, Rodriguez, Ryazantsev, Yu.~Skobelev,
  Soloviev, Willi, Pikuz, Ciardi, \& Fuchs}]{2017SciA....3E0982R}
Revet, G., Chen, S.~N., Bonito, R., {et~al.} 2017, Science Advances, 3, no.11,
  e1700982

\bibitem[{Rodr\'{\i}guez {et~al.}(2018)Rodr\'{\i}guez, Espinosa, \&
  Gil}]{PhysRevE.98.033213}
Rodr\'{\i}guez, R., Espinosa, G., \& Gil, J.~M. 2018, Phys. Rev. E, 98, 033213

\bibitem[{Roe(1981)}]{ROE}
Roe, P. 1981, Journal of Computational Physics, 43, 357

\bibitem[{{Sacco} {et~al.}(2008){Sacco}, {Argiroffi}, {Orlando}, {Maggio},
  {Peres}, \& {Reale}}]{2008A&A...491L..17S}
{Sacco}, G.~G., {Argiroffi}, C., {Orlando}, S., {et~al.} 2008, \aap, 491, L17

\bibitem[{{Sacco} {et~al.}(2010){Sacco}, {Orlando}, {Argiroffi}, {Maggio},
  {Peres}, {Reale}, \& {Curran}}]{2010A&A...522A..55S}
{Sacco}, G.~G., {Orlando}, S., {Argiroffi}, C., {et~al.} 2010, \aap, 522, A55

\bibitem[{{Van Box Som} {et~al.}(2018){Van Box Som}, {Falize}, {Bonnet-Bidaud},
  {Mouchet}, {Busschaert}, \& {Ciardi}}]{2018MNRAS.473.3158V}
{Van Box Som}, L., {Falize}, {\'E}., {Bonnet-Bidaud}, J.~M., {et~al.} 2018,
  \mnras, 473, 3158

\bibitem[{{Zel'dovich} \& {Raizer}(1967)}]{1967pswh.book.....Z}
{Zel'dovich}, Y.~B. \& {Raizer}, Y.~P. 1967, {Physics of shock waves and
  high-temperature hydrodynamic phenomena} (New York: Academic Press,
  1966/1967, edited by Hayes, W.D.; Probstein, Ronald F.)

\end{thebibliography}

\section*{Appendix A: Limits of the FLD approach }
Our radiation hydrodynamics module uses the flux-limited diffusion (FLD) approximation to handle the radiation part of the set of equations \ref{mass}-\ref{eq:energy_cons} (see also Paper I).
The FLD  approach is the most popular method used in radiation hydrodynamics, because of its simplicity, robustness and efficiency. However, it has some limitations. 
In particular, the radiation flux is assumed to be parallel to the gradient of the radiation energy density.
Such an assumption on the direction of the flux, while correct in an optically thick medium, is not true in general.

A first direct consequence is that the radiation force may not be correctly oriented. Another consequence is that the anisotropy of  the radiation field, which arises in non-optically thick media, is not well reproduced in the FLD approach.
For example, in the case of a free-streaming radiation field which encounters a highly opaque region of space, FLD is not able to reproduce a shadow behind the region \citep{Hayes_and_Norman_2003,Gonzalez_et_al_2007}.

An approach more sophisticated than FLD to determine the radiation quantities of the RHD equations is the M1 approximation, which solves the radiation momentum equation to obtain the
radiation flux instead of using the FLD relation (Eq. (23) Paper I). However, for the simulations presented here, we do not expect significant changes of our results if M1 is used instead of FLD. In fact, in the simulated accretion impacts, the radiation force is negligible for the dynamics and the energetics of the accreting system. Furthermore, shadows (not described using the FLD approach) may be produced only in multi-dimensional systems with complex structures alterning optically thick and optically thin parts (e.g., a clumpy accretion column).
 In the case presented in this letter, the portion of the system that is relevant to the determination of the radiation quantities is composed of one optically thin part (the post-shock region) and one optically thick part (the pre-shock region), and the radiation mainly propagates from the optically thin part towards the optically thick part. Thus no shadows are expected in our simulations.

\section*{Appendix B: Role of helium in determining the precursor}
To highlight the role played by helium in the absorption of radiation, we consider a simulation identical to run RHD but using a modified version of the opacity tables, where the helium effects are not taken into account (run RHD-He).

\begin{figure}[!htbp]
    \centering
    \includegraphics[scale =0.45]{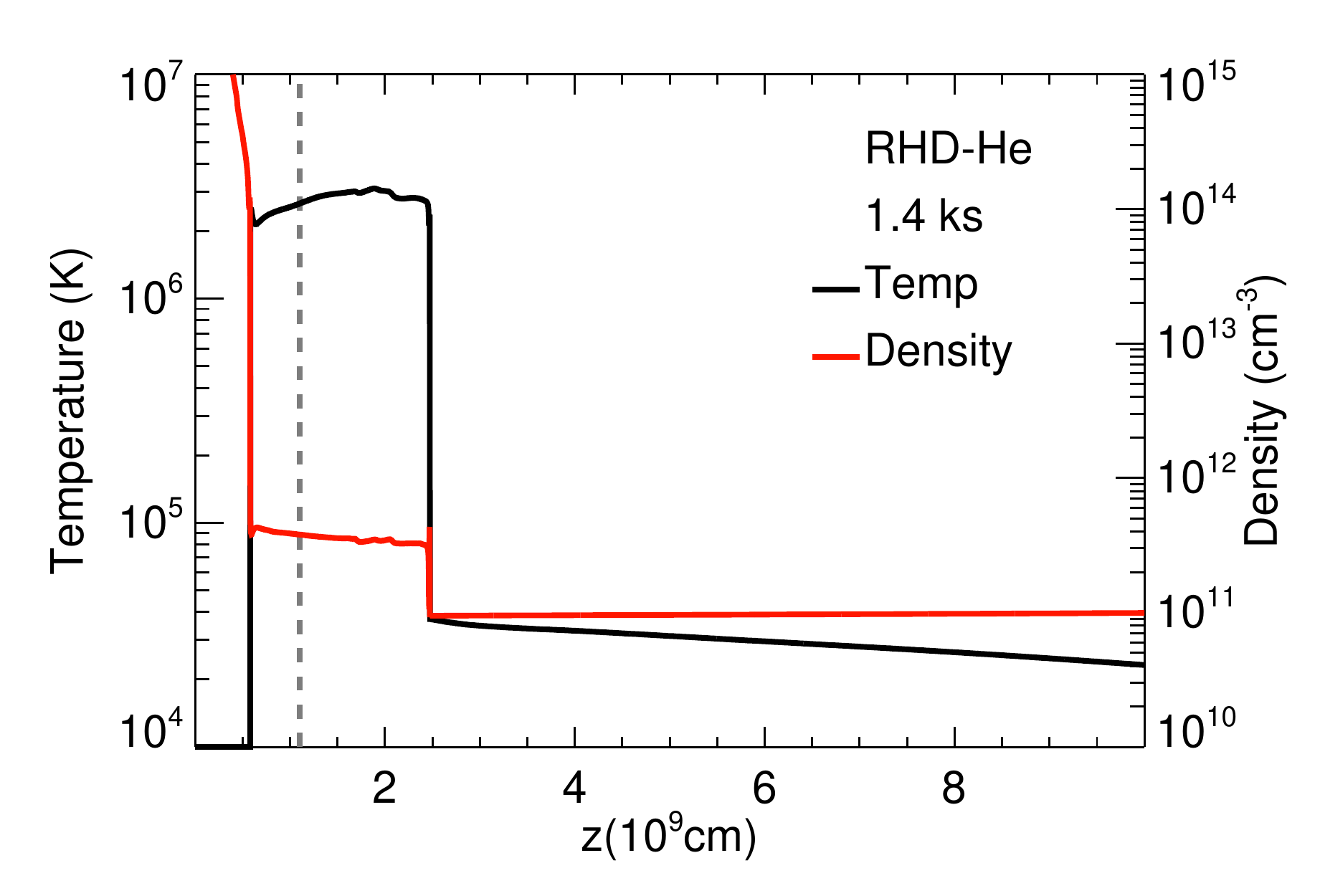}
    \caption{Temperature (black line) and density (red line) profiles for the run RHD-He, during the expanding phase. The grey dotted line represents the initial position of the chromosphere.}
    \label{fig:RHD-He}
\end{figure}

Fig.\ref{fig:RHD-He} is analogous to Fig.~\ref{img:profile} and shows the temperature and density profiles for the run RHD-He. In run RHD-He a precursor forms as in run RHD. However, at odds with run RHD, the downflowing plasma gradually increases its temperature, reaching a maximum temperature of $5\times10^4~$K before being shocked, and no sudden increase of temperature is present. We conclude, therefore, that helium is responsible for the high temperatures of the precursor in run RHD through the peak in $k_P$ at $\log T =4.8$ (See Fig. \ref{img:op}) that determines a sudden increase of absorption.

\end{document}